\documentclass{ws-procs9x6}

\def\etal {{\it et al.}}

\begin{document}

\title{TESTS OF LORENTZ INVARIANCE USING HIGH-ENERGY ASTROPHYSICS OBSERVATIONS}

\author{FLOYD W.\ STECKER}

\address{Astrophysics Science Division, NASA Goddard Sapce Flight Center\\
Greenbelt, MD 20771, USA\\
E-mail: Floyd.W.Stecker@nasa.gov}

\begin{abstract}
High-energy astrophysics observations provide the best possibilities to detect a 
very small violation of Lorentz invariance, such as may be related to the structure of spacetime near the Planck scale. I will discuss the possible signatures of 
Lorentz invariance violation that can be manifested by observing the spectra, polarization, and timing of $\gamma$-rays from active galactic nuclei and $\gamma$-ray bursts. Other sensitive tests are provided by observations of the spectra of ultrahigh-energy cosmic rays and very high-energy neutrinos. I will also discuss a new time-of-flight analysis of observations of GRB 090510 by the Fermi $\gamma$-ray Space Telescope. These results, based on high-energy astrophysical observations, have fundamental implications for spacetime physics and quantum gravity models.
\end{abstract}

\bodymatter

\vskip 20pt
\noindent
{\it Everything that is not forbidden is compulsory.\\
--- Murray Gell-Mann}

\section{Introduction}

Colladay and Kosteleck\'{y}\cite{ck98} proposed an effective field theory framework for quantifying and cataloging the empirical effects of small violations of CPT and Lorentz invariance known
as the Standard-Model Extension (SME). The SME is based on the introduction of small Lorentz and {\cal CPT} violating perturbations in the individual free particle Lagrangians. Coleman and Glashow\cite{co99} have presented a simplified formalism, assuming rotational invariance, wherein particle interactions that violate Lorentz invariance can be modified in terms of the maximum attainable velocities (MAV) of the various particles involved. Thus superluminal particle velocities can be directly related to Lorentz invariance violation. This in turn can lead to the allowance or prohibition of various particle interactions.

\section{Why high-energy astrophysics observations?}
Lorentz invariance implies scale-free spacetime. The group of Lorentz boosts is unbounded.
Very large boosts probe physics at ultrashort distance intervals $\lambda$.
To probe physics at these distance intervals, particularly the nature of space and time, we need to go to ultrahigh energies $E = 1/\lambda$. Cosmic $\gamma$-rays and cosmic rays provide the highest observable energies in the Universe. Physics at the Planck scale of $\sim10^{-35}$ m, e.g., quantum gravity, may involve a fundamental length scale. That, in itself, violates Lorentz invariance. Some frameworks for considering Lorentz invariance violation include effective field theory such as SME, 
deformed special relativity, stochastic spacetime foam, loop quantum gravity,
string inspired models (D-branes), and emergent spacetime theories\cite{dm05}.
I will discuss some particular astrophysical tests of Lorentz invariance violation and limits on MAV.

\section{Limits from photons and electrons}

I follow the well-defined formalism for 
Lorentz invariance breaking discussed in Ref.\ \refcite{co99}.  Within this scenario, 
the MAV of an
electron  need not equal the {\it in vacuo} velocity of light,
i.e., $c_e \ne c_\gamma$. The physical
consequences of this violation of Lorentz invariance depend on the sign of the difference. 
I define the parameter $\delta_{ij} \equiv \delta_{i} - \delta_{j}$ as the
difference between the MAVs of particles $i$ and $j$. 

\subsection{Direct limits}

Direct limits on $\delta_{e \gamma}$ follow from the observation of 2 TeV cosmic-ray
electrons and 50 TeV $\gamma$-rays from the Crab Nebula\cite{co99,sg01}.
These limits are $\delta_{e \gamma} < 1.3 \times 10^{-13}$ and $-\delta_{e \gamma} < 2\times  10^{-16}$. 

\subsection{Threshold for annihilation of $\gamma$-rays
}

If $\delta_{e \gamma} > 0$ the threshold energy for the pair 
production process $\gamma + \gamma \rightarrow e^+ + e^-$ 
is altered 
because the square of the four-momentum becomes
\begin{equation}
2\epsilon E_{\gamma}(1 - \cos \theta) - 2E_{\gamma}^2\delta_{e \gamma} = 4\gamma^2m_{e}^2 >4 m_{e}^2.
\end{equation}
The Mrk 501 $\gamma$-ray spectrum of a strong flare the nearby blazer Mrk 501
above $\sim$20 TeV can 
be understood as a result  of expected intergalactic absorption by pair production. 
It follows that $\delta_{e \gamma} \le 2(m_{e}/E_{\gamma})^2 = 
1.3 \times 10^{-15}$\cite{sg01}.

\subsection{Time of flight for photons from $\gamma$-ray bursts}

The Large Area Telescope aboard the Fermi satellite has placed direct time-of-flight
limits on the energy dependence of photon velocity.\cite{vl13}. This limit rules out the simple
$E/M_{Planck}$ dependence for photon retardation as predicted in some quantum gravity scenarios\cite{el}.

\subsection{Vacuum birefringence} 

Then, using the reported detection 
of polarized soft $\gamma$-ray emission from the $\gamma$-ray burst
GRB041219a that is indicative of an absence of vacuum birefringence, 
together with a very recent improved method for 
estimating the redshift of the burst, I have derived  
strong constraints on the dimension-5 Lorentz invariance violation term
\begin{equation}
{\Delta\cal{L}}_{\gamma} = {{\xi}\over{M_{Pl}}}{n^aF_{ad}n\cdot \partial (n_b \tilde{F}^{bd})}.
\label{deltaL}
\end{equation}
I obtain an upper limit on $|\xi|$ of $2.4 \times 10^{-15}$, corresponding to a constraint on the 
dimension-5 SME coefficient 
$k^{(5)}_{(V)00} \le 4.2 \times 10^{-34}$ GeV$^{-1}$\cite{st11}. It also gives a directional frame
independent constraint in terms of the SME coefficients of
\begin{equation}
|\sum_{jm}{Y_{jm}}(37^{\circ},0^{\circ})k^{(5)}_{(V)jm}| \le 1.2 \times 10^{-34}\ {\rm GeV^{-1}}.
\end{equation}
Other directional constraints from various $\gamma$-ray bursts have recently been derived\cite{ko13}.

\subsection{Synchrotron emission from the Crab nebula} 

Synchrotron $\gamma$-rays from the strong April 2011 Crab Nebula flare were observed by the Large Area Telescope on Fermi up to an energy $\sim$400 MeV~\cite{ab11}.  This places an upper limit on $\delta_{e \gamma}$ of  $6 \times 10^{-20}$\cite{st13}. This equals the limit on the  SME parameter  $-c^e_{TT}$.

\section{Limits from neutrinos
}

The observation of two PeV-scale neutrino events reported by Ice Cube\cite{aa13} can, in principle, 
allow one to place constraints on Lorentz invariance violation in the neutrino sector. This observation implies
an upper limit on $\delta_{\nu e}$ of $\sim3.1 \times 10^{-19}$. Combined with the best limit
on $\delta_{e \gamma}$ given in Section 3.5, this gives an upper limit on $\delta_{\nu \gamma}$ of
$\sim3 \times 10^{-19}$\cite{st13}. This equals the limit on the  SME parameter $-\mathaccent'27 c^{(4)}$

\section{Limits from hadrons
}

Owing to the GZK effect\cite{gr66,za66}, protons with energies above 100 EeV should be attenuated from distances beyond $\sim$100 Mpc because they interact with the CBR photons with a resonant photoproduction of pions\cite{st68}. The threshold energy for this reaction is changed by 
$2\delta_{\pi p}E_{\pi}^2$. This leads to an upper limit 
$\delta_{\pi p} \le 3.23 \times 10^{-24} (\omega/\omega_0)^2$ where
$\omega_0 = kT_{CMB} = 2.35 \times 10^{-4}$ eV with $T_{CMB} = 2.725\pm 0.02$ K\cite{co99}. By comparing the
observed spectrum of ultrahigh energy cosmic rays with the predicted modification from Lorentz invariance violation 
an extremely low upper limit on $\delta_{\pi p}$ of $4.5 \times 10^{-23}$ is obtained\cite{ss09}.

\section{Conclusion}

Presently, we have no positive evidence for modifying special relativity at even the highest energies observed. Theoretical models involving Planck scale physics and quantum gravity need to meet all of the present observational constraints.

\end{document}